\documentclass[a4paper,11pt]{article}
\usepackage{jinstpub} 
\usepackage{lineno}
\usepackage{bm}

\title{\boldmath Study of Bulk Damage of High Dose Gamma Irradiated p-type Silicon Diodes with Various Resistivities}

\author[a,e,1]{I. Zatocilova,\note{Corresponding author.}}
\author[a]{M. Mikestikova,}
\author[a]{V. Latonova,}
\author[a]{J. Kroll,}
\author[a,b]{R. Privara,}
\author[a,c,f]{P. Novotny,}
\author[d]{D.~Dudas,}
\author[a]{and J. Kvasnicka}
\affiliation[a]{Institute of Physics of the Czech Academy of Sciences,\\ Na Slovance 1999/2, 182 00 Prague 8, Czech Republic}

\affiliation[b]{Faculty of Science, Palacky University Olomouc,\\  17. listopadu 1192/12, 779 00 Olomouc, Czech Republic}
\affiliation[c]{Faculty of Mathematics and Physics, Charles University,\\ V Holešovičkách 2, 180 00 Prague 8, Czech Republic}
\affiliation[d]{UJP PRAHA a.s.,\\  Nad Kamínkou 1345, 156 10 Prague – Zbraslav, Czech Republic}
\affiliation[e]{now at Physikalisches Institut, Albert-Ludwigs-Universität Freiburg,\\Hermann-Herder-Straße 3, 79104 Freiburg, Germany}
\affiliation[f]{now at Faculty of Physics, Weizmann Institute of Science, \\ 234 Herzl Street, Rehovot 7610001, Israel}

\emailAdd{iveta.zatocilova@cern.ch}

\abstract{The bulk damage of p-type silicon detectors caused by high doses of gamma irradiation has been studied. The study was carried out on three types of n$^{+}$-in-p silicon diodes with~comparable geometries but different initial resistivities. This allowed to determine how different initial parameters of studied samples influence radiation-induced changes in the measured characteristics. The~diodes were irradiated by a Cobalt-60 gamma source to total ionizing doses ranging from~0.50 up~to~8.28 MGy, and annealed for 80 minutes at 60 °C. The Geant4 toolkit for simulation of~the~passage of particles through matter was used to simulate the deposited energy homogeneity, to~verify the~equal distribution of total deposited energies through all the layers of irradiated samples, and~to~calculate the secondary electron spectra in the irradiation box. The main goal of the study was to characterize the gamma-radiation induced displacement damage by measuring current-voltage characteristics (IV), and the evolution of the full depletion voltage ($V_{\mathrm{FD}}$) with~the~total ionizing dose, by measuring capacitance-voltage characteristics (CV). It has been observed that the bulk leakage current increases linearly with total ionizing dose, and the damage coefficient depends on the initial resistivity of the silicon diode. The effective doping concentration and~therefore $V_{\mathrm{FD}}$ significantly decreases with increasing total ionizing dose, before starting to increase again at a specific dose. We assume that this decrease is caused by the effect of acceptor removal. Another noteworthy observation of this study is that the IV and CV measurements of the gamma irradiated diodes do not reveal any annealing effect.}

\keywords{Particle tracking detectors (Solid-state detectors); Radiation damage to detector materials (solid state); Radiation-hard detectors; Solid state detectors}

\begin{document}
\maketitle
\flushbottom

\section{Introduction}
\label{sec:intro}
Silicon detectors that are used for tracking purposes in large particle accelerator experiments operate in high radiation environments. High particle fluences and ionizing doses introduce defects in~the~silicon substrate, causing bulk damage, as well as in the SiO$_{\mathrm{2}}$ insulating layer and the Si-SiO$_{\mathrm{2}}$ interface, leading to the deterioration of the sensor surface.\par

The primary cause of bulk damage in the silicon detectors, induced by hadrons or gamma rays with higher energies, is the displacement of the Primary Knock-on Atom (PKA) from its lattice site. The~minimum energy required for this process is $\approx$~25 eV. Such a single displacement resulting in~a~pair of a silicon interstitial and a vacancy (Frenkel pair) can be generated by e.g. neutrons or~electrons with an energy above 175 eV and 260 keV respectively. For our investigation, silicon samples were irradiated by Cobalt-60 gamma rays. In this case, the displacement damage is caused primarily due to the interaction of Compton electrons with energies of a few hundreds of~keV, with~the~maximum energy of approximately 1 MeV. It is important to note that production of~clusters of~defects is not possible as the minimum electron energy needed for production of~clusters is $\approx$~8~MeV and~the~maximum recoil energy for the Si PKA by Compton electron is only around 140~eV. Consequently, the damage observed in our silicon samples is exclusively attributed to~the~point defects. \par

The bulk damage of silicon detectors induced by gamma rays from a Cobalt-60 source was previously studied in detail for n-type bulk \cite{a}, revealing an initial drop of the full depletion voltage up to a specific dose of about 2 MGy followed by a subsequent increase of the full depletion voltage up to 9 MGy. This behaviour has been interpreted as a result of the creation of deep acceptor-like defects and donor removal, which lead to a decrease of the initially positive space charge followed by its total compensation at a certain dose, and thereafter to the inversion of the space charge sign. \par

Given that the vast majority of semiconductor detectors recently developed for experiments to~be installed at the High Luminosity LHC use the p-type bulk silicon technology, a detailed study of gamma induced damage of p-type silicon detectors was highly demanded. The first studies performed on p-type silicon samples irradiated by gamma rays up to 3 MGy have shown a~clear decrease of the full depletion voltage and effective doping concentration with increasing total ionizing dose \cite{b}. \par

The presented study brought up a goal to study radiation damage of the high resistivity p-type silicon samples irradiated to high total ionizing doses up to 8.28 MGy using a variety of samples with~different resistivities. \par

\section{Sample Characterisation}
The study was carried out on three types of n$^{+}$-in-p float zone silicon diodes with comparable geometries but different initial resistivities. Diodes were produced by three different manufacturers, Centro Nacional de Microelectrónica (CNM), Hamamatsu Photonics (HPK), and Infineon Technologies (IFX). \par

Initial parameters of the tested samples are listed in table~\ref{tab:i}. The active volume $V$ of the diode was calculated using known active area $A$, that is an area of the diode bounded by a guard ring, and~active thickness $d$ quoted by the manufactureres in diode's specifications. The bulk capacitance of~the~diode $C_{\mathrm{bulk}}$ was estimated as 
\begin{equation}
\begin{aligned}
C_{\mathrm{bulk}}=\varepsilon\frac{A}{d}
\end{aligned}
\end{equation}
where $\varepsilon$ denotes the permittivity of Si (1.06$\cdot$10$^{12}$ F/cm). The initial values of full depletion voltage listed in the table~\ref{tab:i} were obtained from CV characteristics measured for several unirradiated diodes of each type. The full depletion voltage value allows us to determine the bulk resistivity of~the~silicon wafer $\rho$ as
\begin{equation}
\begin{aligned}
\label{rho}
\rho=\frac{d^2}{2\varepsilon \mu V_{\mathrm{FD}}}
\end{aligned}
\end{equation}
where $\mu$ is the hole mobility. Resistivities $\rho$ stated in the table~\ref{tab:i} correspond to the averaged values calculated using formula \ref{rho}. Initial silicon resistivity $\rho$ varies for each type of diode but it is similar for HPK and IFX samples. \par

\begin{table}[htbp]
\centering
\caption{Initial characteristics of CNM, HPK, and IFX diodes.\label{tab:i}}
\smallskip
\begin{tabular}{c|ccc}
\hline
&\textbf{CNM}&\textbf{HPK}& \textbf{IFX}\\
\hline
\textbf{Active thickness $\bm{d}$} & 285 $\mu$m & 290 $\mu$m & 285 $\mu$m\\
\textbf{Active area $\bm{A}$} & 50.17 mm$^{2}$ & 51.55 mm$^{2}$& 49.95 mm$^{2}$\\
\textbf{Active volume $\bm{V}$} & 0.0143 cm$^{3}$ & 0.0149 cm$^{3}$&0.0142 cm$^{3}$\\
\textbf{Calculated $\bm{C_{\mathrm{bulk}}}$} & 19.88 pF & 19.48 pF & 18.79 pF\\
\textbf{Measured $\bm{V_{\mathrm{FD}}}$} & (36.9 $\pm$ 8.3) V & (273.4 $\pm$ 10.7) V & (283.6 $\pm$ 12.0) V\\
\textbf{Calculated resistivity} $\bm{\rho}$ & (23.975 $\pm$ 4.012) k$\Omega \cdot$cm & (3.301 $\pm$ 0.129) k$\Omega \cdot$cm & (3.077 $\pm$ 0.130) k$\Omega \cdot$cm\\
\hline
\end{tabular}
\end{table}

\section{Irradiation and Geant4 Simulation}
The diodes were irradiated by the Cobalt-60 gamma radionuclide source installed at UJP PRAHA~a.~s.~\cite{c} up to the maximum total ionizing dose (TID) of 8.28 MGy.  \par

During the irradiation, diodes were placed in a custom-made charged-particle equilibrium (CPE) box consisting of an outer 1.5mm-thick Pb layer and an inner layer of 1mm-thick Al layer, following the recommendations of ESCC \cite{d}. While using the CPE box during the irradiation, the dose enhancement from low-energy scattered radiation is minimised by producing electron equilibrium and a uniform distribution of energy deposited in the individual samples is ensured. In~the~CPE box, up to 5 silicon diodes were stacked on top of each other. \par

The dose rate was 160-190 Gy/min in silicon and the dose rate uncertainty was estimated to~be less than 5 \% within the box. The in-silicon dose rate was calculated from the in-air dose rate measured by a calibrated ionizing chamber under the same conditions as the tested samples. The~stated homogeneity of the dose rate field was ensured by using a lead collimator (field homogenizer) in~the~shape of a truncated cone. The cooling of the samples during irradiation was ensured by~using a fan for air ventilation which prevented the temperature from exceeding 35~°C during the~entire irradiation. Afterwards, the irradiated samples were immediately stored in~a~refrigerator at temperatures below -20 °C to prevent any uncontrolled post-irradiation annealing. \par

\begin{figure}[htbp]
\centering
\includegraphics[width=.5\textwidth]{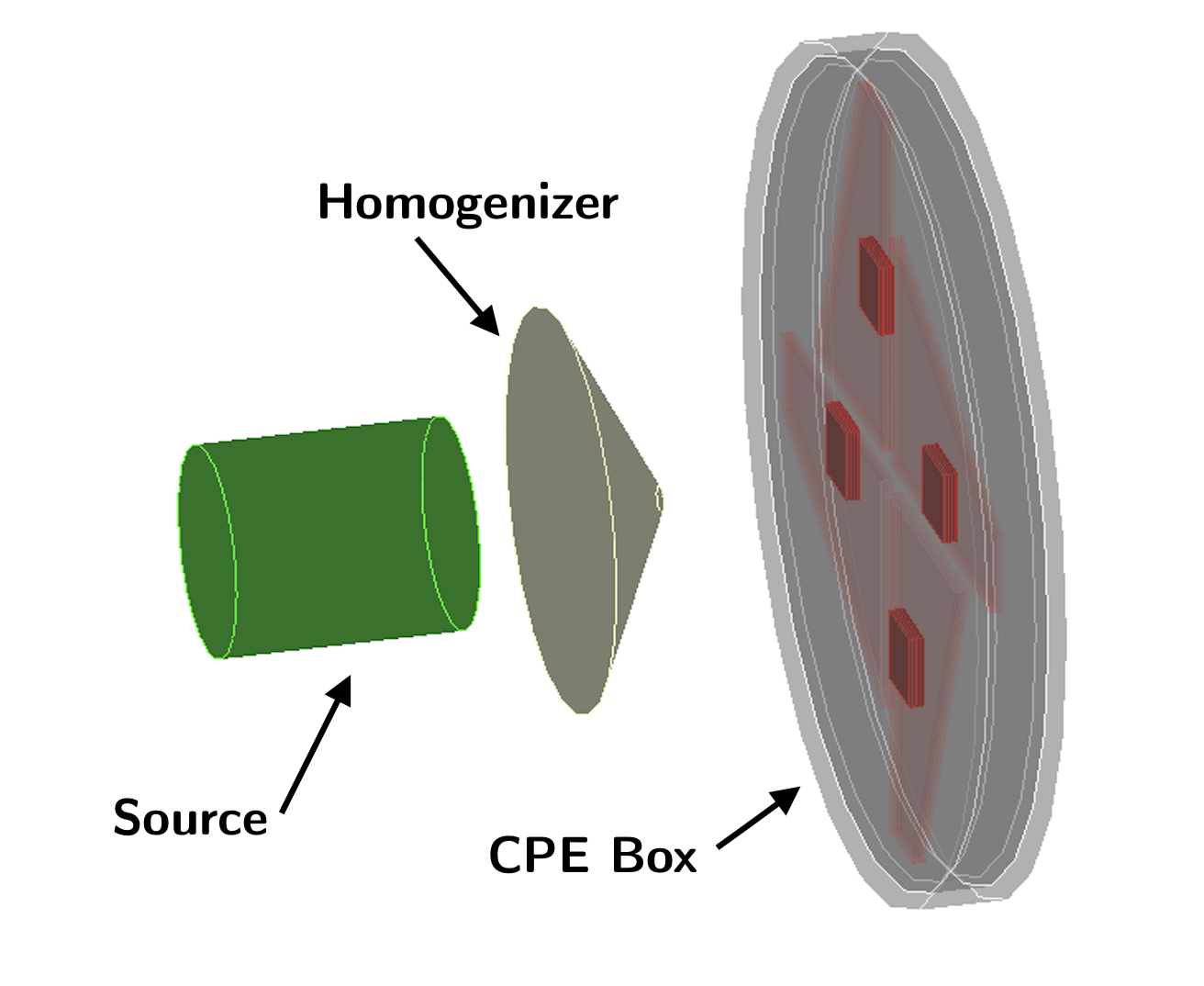}
\caption{Schematic layout of the irradiation setup geometry used in the Geant4 simulation.\label{CPE}}
\end{figure}

The passage of Cobalt-60 gamma ray with an average energy of 1.25 MeV through the silicon samples and surrounding materials causes ionization and production of secondary electrons mainly via Compton scattering. The secondary electrons generated in the shielding material (Pb and~Al) of~the~CPE box, that surrounds the silicon samples, are slowed down on their way through the~shielding before they reach the inner space of the box and cause displacement damage in~the~silicon samples. \par

The Monte Carlo transport code Geant4 was used to simulate the deposited energy homogeneity, to verify the equal distribution of total deposited energy through all the layers of irradiated samples, and to calculate the secondary electron spectra in the CPE box. In the simulation, monoenergetic photons with the energy of 1.25 MeV were emitted isotropically from the source and sent through the~lead collimator, as well as through the aluminium and lead layers of the CPE box. The~schematic of the irradiation setup geometry implemented in~the~simulation is shown in figure~\ref{CPE}. \par

The simulation clearly demonstrates that usage of~the~CPE box for gamma irradiation of~studied samples ensures a sufficiently homogeneous distribution of~the~deposited energy which varies within $\pm$ 5 \% in the irradiation area. The results of the simulation showed that the maximum energy of~the~secondary electrons leaving the inner aluminium layer of the CPE box is $\approx$ 1 MeV. The~simulation also demonstrated a reasonably homogeneous distribution of the deposited energy in all the layers of silicon during irradiation in the CPE box.\par

\section{Experimental Methods and Results}
To evaluate the effects of bulk radiation damage induced by gamma rays, all the samples were at~first measured before the irradiation to determine their initial parameters. After gamma irradiation, the~samples were tested both before and after their annealing for 80 minutes at 60 °C. The measured electrical parameters were the leakage current and bulk capacitance of the diodes as a function of~the~bias voltage (IV and CV characteristics). \par 

A great care was taken to properly determine the leakage current contributing only to the active volume of the diode by subtracting the parasitic currents contributed by the diode surface. The~total current $I_{\mathrm{tot}}$ is defined as the sum of the surface current and the bulk leakage current $I_{\mathrm{leak}}$. Only the~bulk current flowing through the active volume of the diode, which is bounded by the guard ring, will be noted as the diode leakage current in the following text. The measurement is performed by~applying negative bias voltage to the backside of the diode, however the absolute value of bias voltage is used for all discussion in the text of this paper. \par

The measurements of IV and CV characteristics took place in a controlled environment at~the~temperature of (20 $\pm$ 2) °C and relative humidity ($RH$) lower than 10 \%. In order to be able to compare directly the obtained IV characteristics, the measured bulk leakage currents were normalized to the temperature of +20 °C. \par

\subsection{Leakage current}
\begin{figure}[htbp]
\centering
\includegraphics[width=0.8\textwidth]{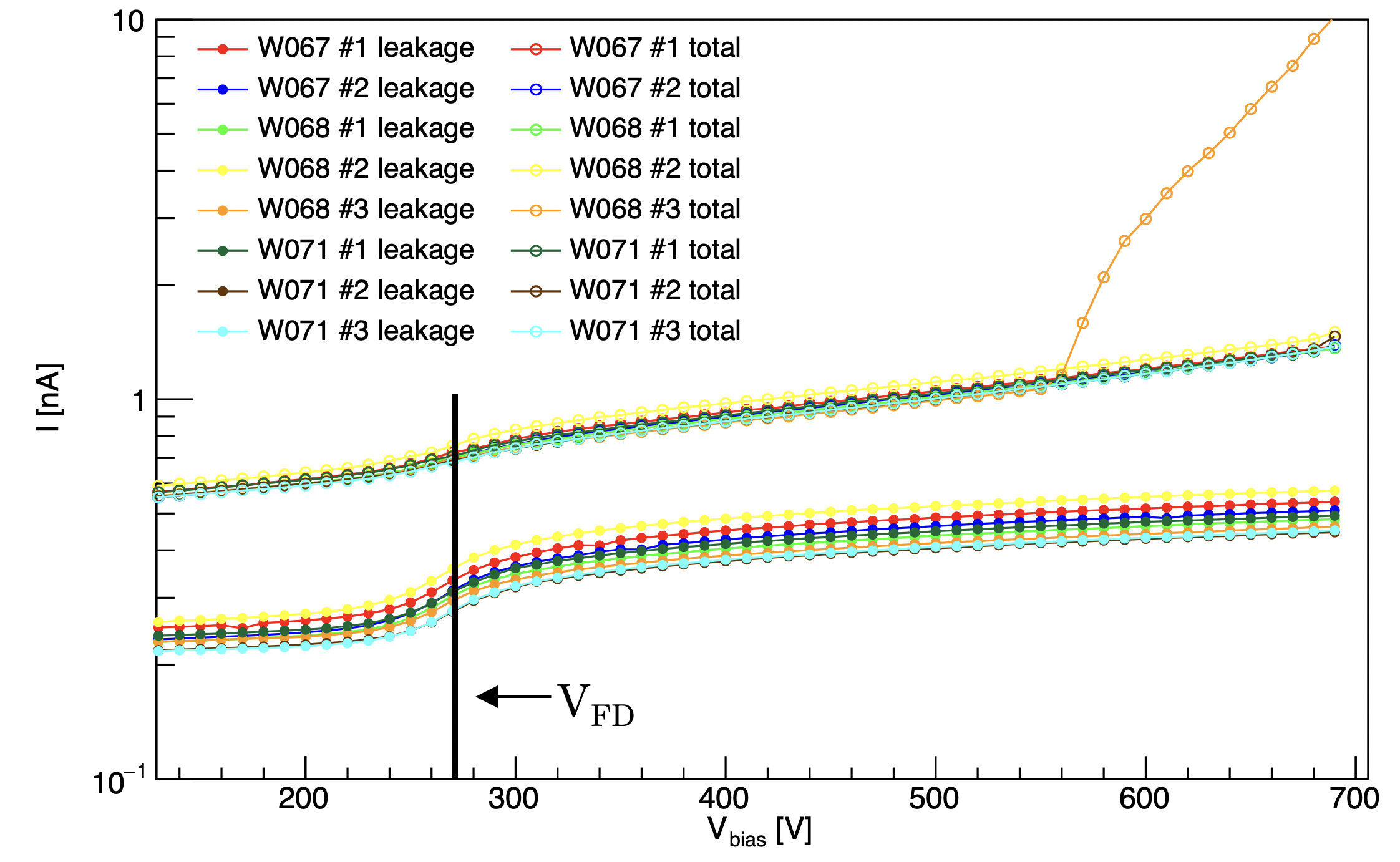}
\caption{Dependence of the bulk leakage current $I_{\mathrm{leak}}$ (full circles) and the total reverse current $I_{\mathrm{tot}}$ (open circles) of unirradiated HPK diodes on the bias voltage $V_{\mathrm{bias}}$. \label{IVunirrad}}
\end{figure}

To illustrate the IV characteristics measured on diodes prior to irradiation, figure~\ref{IVunirrad} displays the~total and leakage currents of 8 unirradiated HPK diodes. The CNM and IFX diodes that are not shown in this plot exhibited equivalent behavior. The IV characteristics of diodes reveal uniform behavior up to 700 V, the total current is more than 2 times greater than the bulk leakage current. While the~saturation of the bulk leakage current at the full depletion voltage $V_{\mathrm{FD}}$ is clearly visible, the total currents do not exhibit the same behavior. Instead they increase with the increasing bias voltage $V_{\mathrm{bias}}$. This observation indicates that for the study of bulk damage induced by gamma rays it is very important to separate surface and bulk currents. \par

\begin{figure}[htbp]
\centering
\includegraphics[width=0.9\textwidth]{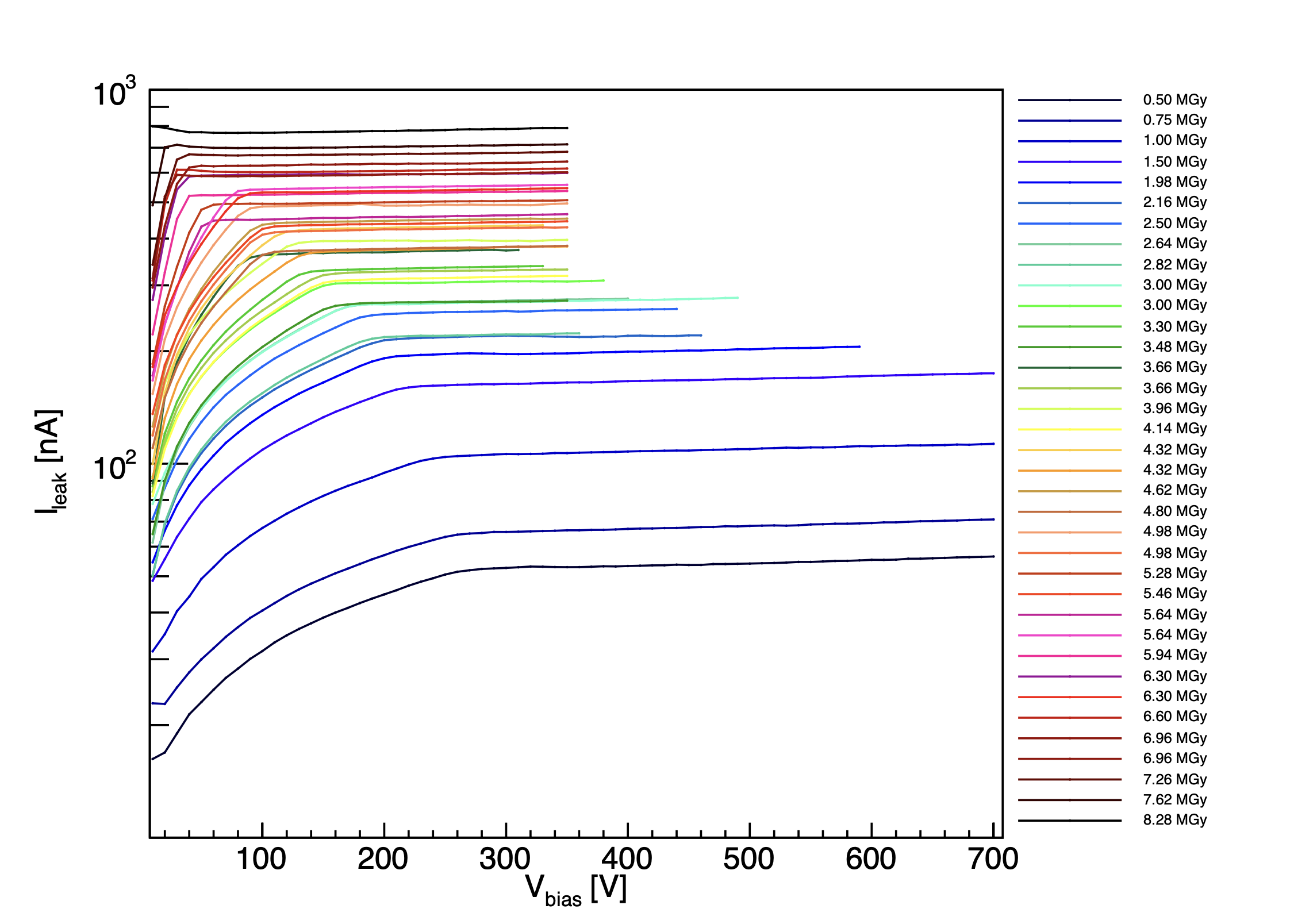}
\caption{Bulk leakage current as a function of bias voltage measured for HPK diodes irradiated to various TIDs ranging from 0.5 MGy to 8.28 MGy. All displayed measurements were performed after annealing. \label{IVirrad}}
\end{figure}

Figure~\ref{IVirrad} shows the IV characteristics of HPK diodes irradiated to various TIDs after their annealing. The CNM and IFX diodes that are not shown in this plot exhibited similar behavior. A general behavior of the leakage current as a function of the bias voltage is equivalent to the IV characteristics of unirradiated diodes - the leakage current is increasing with increasing bias voltage until the full depletion is reached, then it remains constant. An increase of the leakage current of fully depleted irradiated diodes with increasing TID is observed for all 3 types of diodes. A comparison of the bulk leakage current measurements of gamma irradiated n$^{+}$-in-p diodes performed before and~after their annealing indicates that the leakage current remains unchanged with the application of annealing. The IV characteristics measured before annealing are not shown in the plot.\par

\begin{figure}[htbp]
\centering
\includegraphics[width=.8\textwidth]{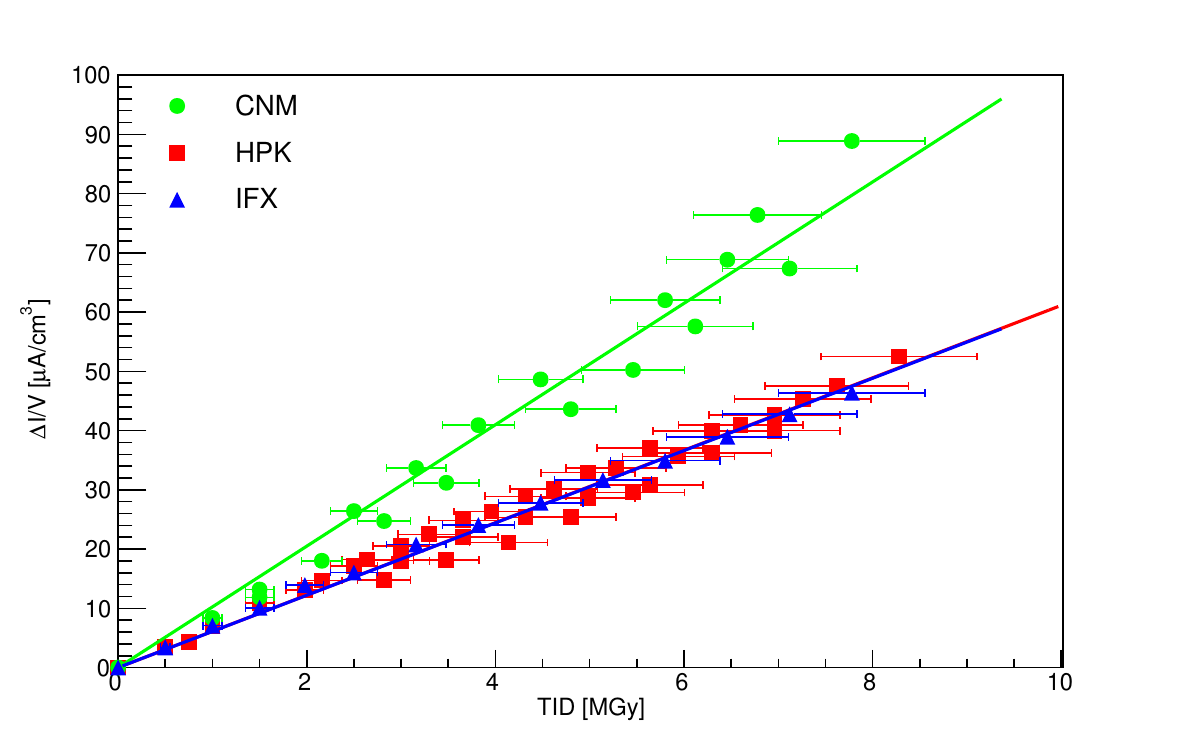}
\caption{Radiation-induced leakage current increase $\Delta I/V$ as a function of the TID of fully depleted CNM, HPK and IFX diodes. Measured currents are normalized to 20 °C. Diodes are annealed at 60 °C for~80~minutes. \label{IV}}
\end{figure}

The radiation-induced current per active volume of the diode $\Delta I/V$ is plotted as a function of the~TID in figure~\ref{IV}. The quantity $\Delta I$ is defined as the difference between the leakage current of~the~fully depleted diode measured after and before its irradiation, measured at 60 V for CNM diodes and~at~300~V for HPK and IFX diodes. $V$ stands for the volume contributing to the current. The leakage current is normalized to 20 °C and it is measured after the annealing.\par

In the studied range up to 8 MGy the values of the leakage current obtained for all types of~n$^{+}$-in-p silicon diodes show a linear increase with the TID. For the CNM diode, which has the~highest bulk resistivity, the change  of the leakage current with the delivered TID is faster than for~the~HPK and IFX diodes, which have much lower resistivities. We can express the relation between~$\Delta I/V$ and TID by~the~formula
\begin{equation}
\begin{aligned}
\frac{\Delta I}{V} = a_{\mathrm{\gamma}} \cdot TID
\end{aligned}
\end{equation}
where $a_{\mathrm{\gamma}}$ is the current-related damage coefficient. \par

The obtained values of damage coefficient $a_{\mathrm{\gamma}}$ for each diode type are listed in table~\ref{tab:ii}. \par

\begin{table}[htbp]
\centering
\caption{The averaged values of the damage coefficient $a_{\mathrm{\gamma}}$, initial silicon resistivity $\rho$ and initial full depletion voltage $V_{\mathrm{FD}}$ measured for CNM, HPK, and IFX diodes.\label{tab:ii}}
\smallskip
\begin{tabular}{cccc}
\hline
& $a_{\mathrm{\gamma}}$ [A$\cdot$cm$^{-3} \cdot$MGy$^{-1}$] & $\rho$ [k$\Omega \cdot$cm]& $V_{\mathrm{FD}}$ [V]\\
\hline
CNM & (10.20 $\pm$ 0.30) $\cdot$10$^{-6}$& (23.975 $\pm$ 4.012)  & (36.9 $\pm$ 8.3)\\
HPK & (6.49 $\pm$ 0.09) $\cdot$10$^{-6}$ & (3.301 $\pm$ 0.129) & (273.4 $\pm$ 10.7)\\
IFX & (6.33 $\pm$ 0.08) $\cdot$10$^{-6}$ & (3.077 $\pm$ 0.130) &(283.6 $\pm$ 12.0) \\
\hline
\end{tabular}
\end{table}

The measured values of $a_{\mathrm{\gamma}}$ indicate that diodes with similar initial resistivities and full depletion voltages (HPK and IFX) show the same level of radiation damage. On the other hand, CNM diodes with a higher initial resistivity and lower full depletion voltage have a higher damage coefficient $a_{\mathrm{\gamma}}$. The TID of 1 MGy  causes the bulk current change of $\approx$ 10.2 $\mu$A/cm$^{3}$ in CNM diode, and~$\approx$~6.4~$\mu$A/cm$^{3}$ in HPK and IFX diodes. \par

\subsection{Bulk Capacitance and Full Depletion Voltage}
The bulk radiation-induced defects lead to a change of the effective space charge $N_{\mathrm{eff}}$ that is reflected in a change of the full depletion voltage $V_{\mathrm{FD}}$ of the silicon diode. The $V_{\mathrm{FD}}$ is given as
\begin{equation}
\begin{aligned}
\label{xy}
V_{\mathrm{FD}} = \frac{q|N{\mathrm{eff}}|d^2}{2\varepsilon\varepsilon_0}
\end{aligned}
\end{equation}
where $d$ is the active thickness of the diode, $q$ is the elementary charge, $\varepsilon$ is the relative permittivity of silicon, and $\varepsilon_{\mathrm{0}}$ is the vacuum permittivity. This equation assumes a constant space charge distribution over the volume of the diode. Determination of $N_{\mathrm{eff}}$ from \eqref{xy} might be afflicted with~systematic errors, but in the case of gamma irradiation we can assume that the silicon bulk is not heavily defected, and the space charge is identical to the free carrier concentration in thermal equilibrium.  \par

The full depletion voltage in this study was determined from the CV characteristics as~the~value of bias voltage for which the linear increase of 1/$C_\mathrm{bulk}^2$ dependence reaches its plateau. The~bulk capacitance was measured by using the testing signal with the amplitude of 2 V and frequency of~1~kHz (for~unirradiated diodes) and 100 kHz (for irradiated diodes). In order to determine the~bulk capacitance values precisely, the guard ring was grounded also during the CV measurements.  The~values of~$C_\mathrm{bulk}$ obtained for fully depleted unirradiated diodes are to 19.30 pF, 19.52 pF, and~18.47~pF for~HPK, CNM, and IFX, respectively. These values are in very good agreement with~calculated values listed in table \ref{tab:i} and prove a good quality of the measuring setup. \par

\begin{figure}[htbp]
\centering
\includegraphics[width=0.8\textwidth]{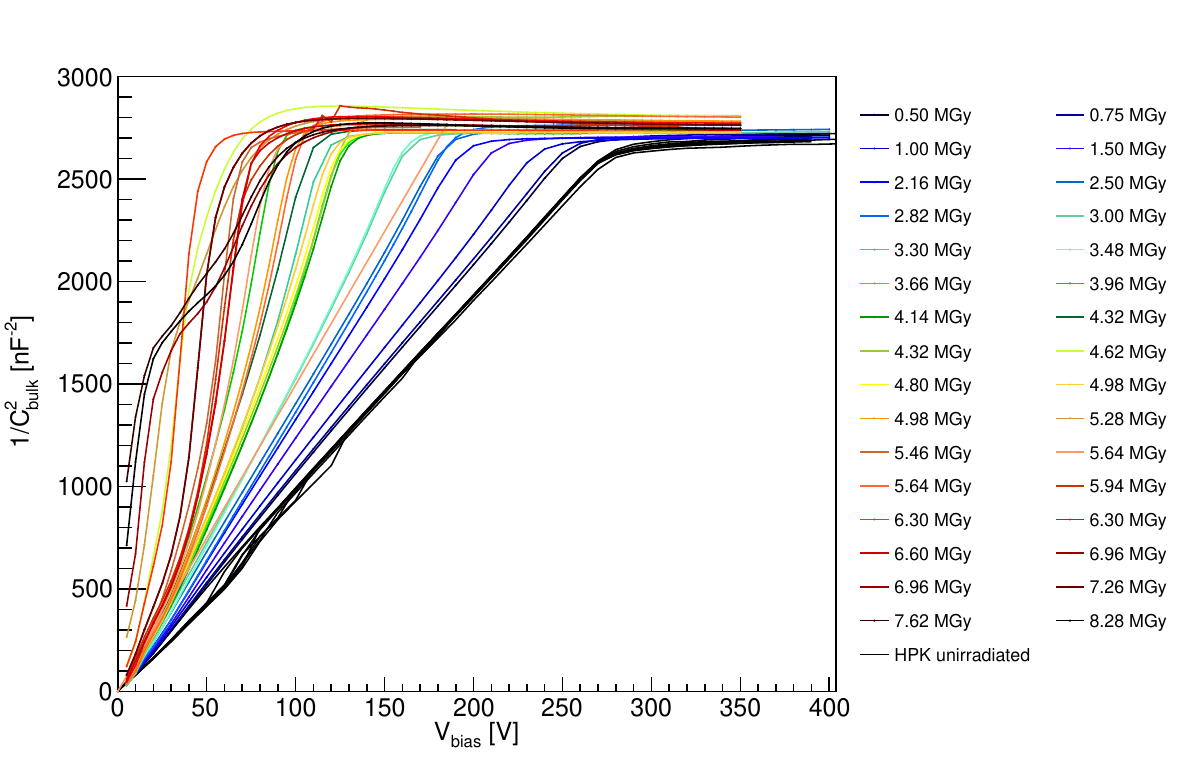}
\caption{CV characteristics of HPK diodes irradiated to various TIDs and measured after annealing at~60~°C for~80 minutes. Data obtained for unirradiated HPK diodes (black lines) is shown for comparison. \label{CV_HPK}}
\end{figure}

Dependencies of 1/$C_\mathrm{bulk}^2$ on the applied bias voltage measured for the HPK diodes irradiated to~various TIDs after their annealing are plotted in figure \ref{CV_HPK}.  The CNM and IFX diodes that are not shown in this plot exhibited similar behavior. The plot also includes an equivalent dependence obtained for~the~unirradiated diode as a reference. The evolution of the bulk capacitance $C_\mathrm{bulk}$  with~the~TID is clearly visible. \par

\begin{figure}[htbp]
\centering
\includegraphics[width=1\textwidth]{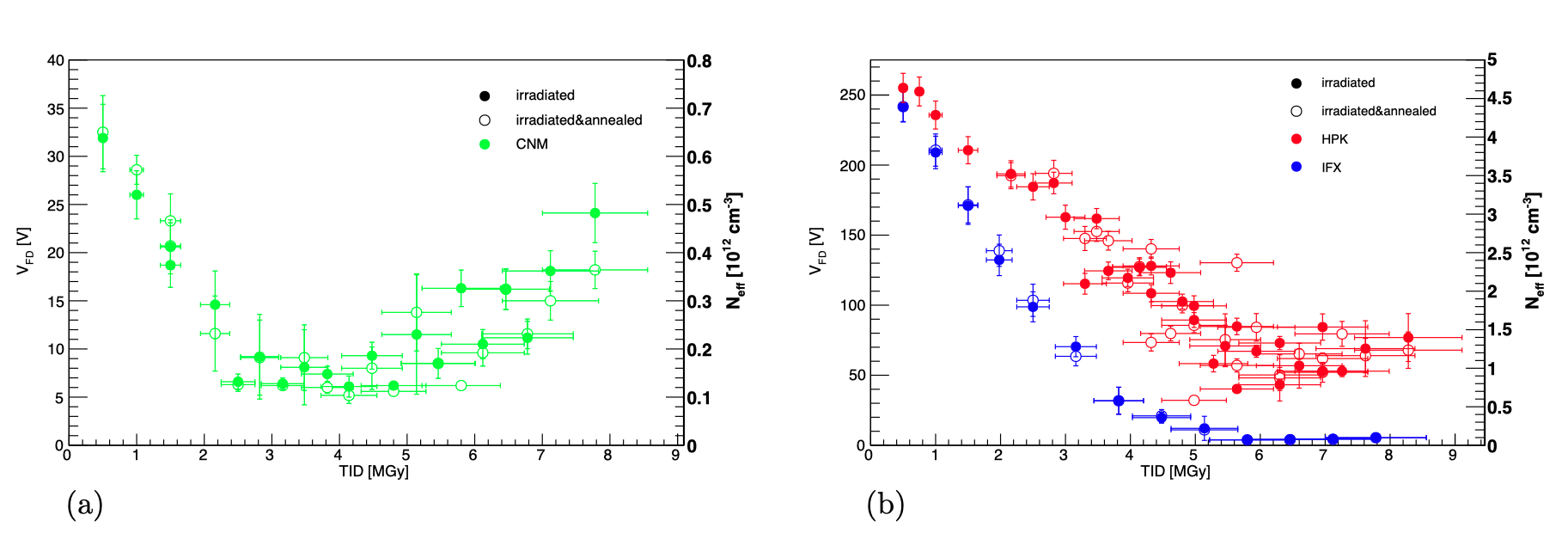}
\caption{(a) Dependence of the $V_{\mathrm{FD}}$ of irradiated CNM diodes on the delivered TID, measured both before (full marks) and after the standard annealing (open marks).\label{FDV_CNM} (b) Dependence of the $V_{\mathrm{FD}}$ of irradiated HPK and IFX diodes on the delivered TID, measured both before (full marks) and after annealing (open marks). \label{FDV_HPK_IFX}}
\end{figure}

The evolution of the full depletion voltage with the increasing TID is shown in figure \ref{FDV_CNM} for~CNM and HPK/IFX diodes. 
The full depletion voltage $V_{\mathrm{FD}}$, and thus also the effective doping concentration $N_{\mathrm{eff}}$, of p-type silicon diodes measured both before and after their annealing significantly decreases with~the~increasing TID to a certain minimum value, and then increases again with~increasing dose. The~TID value, for which the $V_{\mathrm{FD}}$ reaches its minimum, reflects the initial resistivity of the diode.  For~the~CNM diodes, which have the highest initial resistivity of $\approx$ 24 k$\Omega \cdot$cm, the~minimum value of~$V_{\mathrm{FD}}$ is reached for the TID of 4.14 MGy. The HPK and IFX diodes, which have a comparable initial resistivity of~$\approx$~3.1~-~3.3~k$\Omega \cdot$cm, reach the minimal $V_{\mathrm{FD}}$ at the TID close to~6~MGy. For IFX diodes the measured $V_{\mathrm{FD}}$ is not increasing in the measured interval of TID. \par

No systematic effect of annealing on the obtained $V_{\mathrm{FD}}$ is observed. Differences in $V_{\mathrm{FD}}$ values obtained before and after annealing are negligible in comparison to standard deviations of~determined $V_{\mathrm{FD}}$ values. \par

\section{Conclusions}
The bulk damage of high resistivity p-type silicon diodes caused by high doses of gamma irradiation has been studied. The study was carried out on three types of n$^{+}$-in-p diodes with different initial resistivities but comparable geometries. The diodes were irradiated by a  Cobalt-60 gamma source to~various TIDs ranging from 0.50 to 8.28 MGy, and annealed for 80 minutes at 60 °C. The main goal of the presented research was to study the gamma radiation-induced displacement damage by analysing obtained current-voltage characteristics and to understand the evolution of the full depletion voltage ($V_{\mathrm{FD}}$) with TID by measuring capacitance-voltage characteristics of the tested diodes. \par

It was observed that the bulk leakage current increases linearly with the TID, and~the~damage coefficient depends on the initial resistivity of the silicon diode. The damage coefficients of~the~diodes with initial silicon resistivities of~$\approx$~24, 3.3, and~3.1~k$\Omega \cdot$cm were determined to be 10.20$\cdot$10$^{-6}$, 6.49$\cdot$10$^{-6}$, and 6.33$\cdot$10$^{-6}$ A$\cdot$cm$^{-3} \cdot$MGy$^{-1}$, respectively.  \par

The effective doping concentration ($N_{\mathrm{eff}}$), and therefore also the $V_{\mathrm{FD}}$, significantly decreases with the increasing TID, before it starts increasing at a specific TID value. A similar effect was observed in p-type silicon samples irradiated by hadrons \cite{f}, where the initial decrease of~the~full depletion voltage was explained by electrically active boron dopants being deactivated while~the~sample is irradiated to lower fluences. After being irradiated by hadrons to higher fluences, the~samples exhibit a different behavior when affected by other radiation-induced effects. Our study reveals that diodes with higher initial resistivity reach the minimum value of $V_{\mathrm{FD}}$ at~a~lower TID compared to~diodes with a lower initial resistivity. \par 

Another important conclusion of this study is that annealing for 80 minutes at 60 °C, typically used for hadron-irradiated silicon sensors, has no effect on the gamma radiation-induced damage in~p-type silicon. There are no changes in the full depletion voltage or leakage current values observed after such an annealing. This annealing behavior of gamma-irradiated n$^{+}$-in-p silicon diodes is quite different from that for hadron-irradiated diodes. The fact that neither the $V_{\mathrm{FD}}$ nor~the~leakage current changes with annealing could originate from predominantly immobile defects. \par


\acknowledgments

This research was funded by the Ministry of Education, Youth and Sports of the Czech Republic grant number LTT17018 Inter-Excellence and LM2018104 CERN-CZ.


\end{document}